\documentclass[prl,twocolumn,showpacs,preprintnumbers,amsmath,amssymb]{revtex4}

\usepackage{graphicx}
\usepackage{dcolumn}
\usepackage{bm}

\begin{document}


\title{Experimental Bell Inequality Violation with an Atom and a Photon}

\author{D. L. Moehring}
 \email{dmoehrin@umich.edu}
\author{M. J. Madsen}
\author{B. B. Blinov}
\author{C. Monroe}
\affiliation{FOCUS Center and Department of Physics, University of Michigan, Ann Arbor, Michigan 48109-1120}

\date{\today}

\begin{abstract}
We report the measurement of a Bell inequality violation with a single atom and a single photon prepared in a probabilistic entangled state.  This is the first demonstration of such a violation with particles of different species.  The entanglement characterization of this hybrid system may also be useful in quantum information applications.
\end{abstract}


\maketitle

The famous 1935 Einstein-Podolsky-Rosen thought experiment showed how measurements of certain entangled quantum systems require a nonlocal description of nature \cite{einstein:1935}, thus leading to the suggestion that quantum mechanics is incomplete.  However, starting in 1965, Bell and others discovered that certain measured correlations between multiple systems, averaged over many identical trials, must obey particular inequalities for \textit{any} local (hidden-variable) theory to apply \cite{bell:1965,clauser:1969}.  Experiments showing violations of these Bell inequalities followed shortly thereafter involving entangled photon pairs \cite{freedman:1972, aspect:1982a, aspect:1982b, oh:1988, weihs:1998}, low-energy protons \cite{lamehi:1976}, neutral kaons \cite{bramon:1999}, and more recently trapped atomic ions \cite{rowe:2001} and individual neutrons \cite{hasegawa:2003}.  There is continued vigorous interest in Bell inequality measurements in these and other systems not only to test the non-locality of quantum mechanics in a variety of contexts, but also because of the close connection between these measurements and quantum entanglement, which is a key resource in the field of quantum information science \cite{nielsen, brukner:2004}.  Furthermore, there is interest in extending current systems to perform a Bell inequality measurement that simultaneously closes various ``loopholes'' in the interpretation of results \cite{kwiat:1994, huelga:1995, fry:1995, simon:2003}.

In this letter, we report the measurement of a Bell inequality violation in the system of a single trapped atom entangled with a single photon, representing the first test of such a violation in a hybrid system.  While this experiment closes neither the locality \cite{aspect:1982b, weihs:1998} nor detection \cite{rowe:2001} loophole, the atom-photon system has future promise to close both loopholes simultaneously \cite{simon:2003}.  In addition, the entanglement of a single atom and a single photon is of great interest in its own right, with known applications in quantum information science including quantum repeater circuits \cite{briegel:1998, dur:1999}, and scalable ion trap quantum computing \cite{duan:2004}.

The form of Bell inequality violated here was first proposed by Clauser, Horne, Shimony, and Holt (CHSH) \cite{clauser:1969}.  Their extension of Bell's original work accommodates non-ideal (experimentally realizable) systems and requires (i) the repeated creation of identical entangled pairs of two-level systems (qubits), (ii) independent rotations of the two qubits, and (iii) measurement of the qubits, each of which has two possible outcomes, $|0\rangle_i$ or $|1\rangle_i$, where $i$ refers to qubit A or B.  The inequality is based on the statistical outcome of correlation function measurements \cite{correlation:footnote} defined as,
\begin{eqnarray}
q(\theta_A,\theta_B)=f_{00}(\theta_A,\theta_B)+f_{11}(\theta_A,\theta_B)\nonumber\\
-f_{10}(\theta_A,\theta_B)-f_{01}(\theta_A,\theta_B),\label{eq:corr}
\end{eqnarray} 
where $f_{\alpha\beta}(\theta_A,\theta_B)$ is the fraction of the total events where particle A was in state $|\alpha\rangle_A$ and particle B was in state $|\beta\rangle_B$ following rotations by polar angles $\theta_A$ and $\theta_B$ on the Bloch sphere.  CHSH show that all local hidden-variable theories must obey the inequality 
\begin{eqnarray}
B(\theta_{A1},\theta_{A2};\theta_{B1},\theta_{B2})\equiv |q(\theta_{A2},\theta_{B2})-q(\theta_{A1},\theta_{B2})|\nonumber\\
+|q(\theta_{A2},\theta_{B1})+q(\theta_{A1},\theta_{B1})| \leq 2.\label{eq:ineq}
\end{eqnarray} 
According to quantum mechanics, however, this inequality can be violated for certain states and measurements.  For instance, quantum theory predicts the state $|\Psi_{i}\rangle=(|0\rangle_A |0\rangle_B+|1\rangle_A |1\rangle_B)/\sqrt{2}$ has a correlation function $q(\theta_A,\theta_B)=\text{cos}(\theta_A-\theta_B)$.  This results in a maximum violation of Eq.~(\ref{eq:ineq}) for certain settings; for example, $B(0,\pi/2;\pi/4,3\pi/4)=2\sqrt{2}$.

The experiment follows the same three steps as above, with a single photon qubit composed of its two orthogonal polarization directions, and a single atom qubit stored in the ground state hyperfine (spin) levels of a trapped $^{111}$Cd$^+$ ion.  The ion is confined in a quadrupole rf ion trap of characteristic size $\sim$0.7 mm and localized to under 1 $\mu$m through Doppler laser cooling.  As described in Ref. \cite{blinov:2004}, the atom and photon are probabilistically entangled following the spontaneous emission of a photon from an excited state of the atom to multiple ground states.  The entanglement is probabilistic due to the small acceptance angle and transmission loss of the photon collection optics ($\sim$1\%), low quantum efficiency of the photon detectors ($\sim$20\%), and a restriction of the excitation probability to $\sim$10\% in order to suppress multiple excitations.  

\begin{figure}
\includegraphics[width=0.8\columnwidth,keepaspectratio]{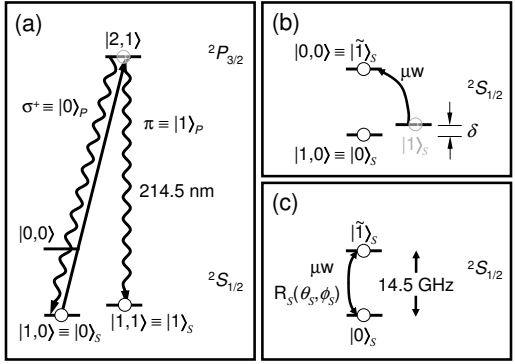}
\caption{\label{fig:f1}The experimental procedure.  (a) The atom is weakly excited with a 50 ns $\sigma^+$-polarized laser pulse at 214.5 nm, resulting in spontaneous emission to either the $|0\rangle_S$ state while emitting a $|0\rangle_P$ polarized photon or to the $|1\rangle_S$ state while emitting a $|1\rangle_P$ polarized photon.  (b) A microwave ($\mu$w) pulse resonant with the $|1,1\rangle\leftrightarrow|0,0\rangle$ transition coherently transfers the population in the $|1\rangle_S$ qubit state to the $|0,0\rangle\equiv|\tilde{1}\rangle_S$ qubit state.  A Zeeman splitting between the $^2S_{1/2}$ $F=1$ levels of $\delta\approx2\pi(1.0 \text{ MHz})$ is provided by the magnetic field defining the quantization axis.  (c) A second microwave pulse, the qubit rotation pulse, R$_S(\theta_S,\phi_S$), prepares the atomic qubit for measurement in any basis.}
\end{figure}

As seen in Figure~\ref{fig:f1}(a), the atom is excited from the $^2S_{1/2}$ $|1,0\rangle$ ground state to the $^2P_{3/2}$ $|2,1\rangle$ state via a 50 ns resonant optical excitation pulse.  The atom then spontaneously decays to either $^2S_{1/2}$ $|1,0\rangle\equiv|0\rangle_S$ while emitting a $\sigma^+$-polarized photon, or $^2S_{1/2}$ $|1,1\rangle\equiv|1\rangle_S$ while emitting a $\pi$-polarized photon.  Here, the quantum numbers $|F,m_F\rangle$ describe the total angular momentum $F$, and its projection along the quantization axis $m_F$, provided by a magnetic field of  $|\textbf{B}|\approx$ 0.7 G.  Along the direction of the collection optics, the polarizations of the $\sigma^+$ photons ($|0\rangle_P$) and the $\pi$ photons ($|1\rangle_P$) are orthogonal \cite{blinov:2004}.  

\begin{figure}
\includegraphics[width=0.8\columnwidth,keepaspectratio]{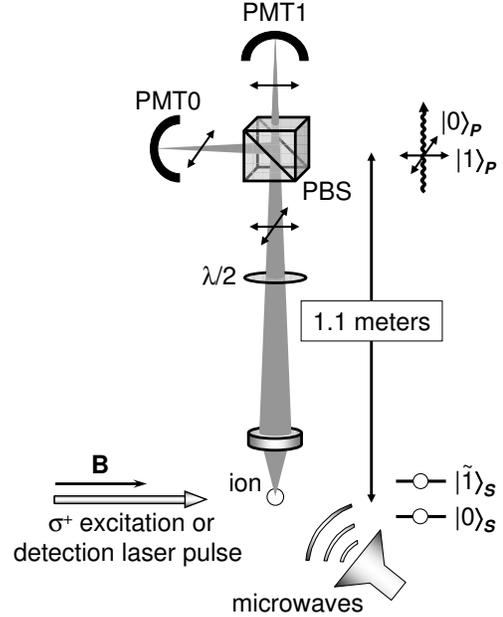}
\caption{\label{fig:f2}The experimental apparatus.  Following excitation of the atom from a 50 ns laser pulse, the scattered photons are collected by an imaging lens, directed to a polarizing beam splitter (PBS), and registered by one of two photon-counting photomultiplier tubes (PMTs).  A $\lambda/2$ waveplate is used to rotate the photon polarization for photonic qubit measurements in different bases, and the microwaves are similarly used on the ion to drive coherent transitions between the atomic qubit hyperfine ground states at a frequency near 14.5 GHz.  Following application of the microwaves, a 125 $\mu$s laser pulse is directed on the ion for atomic fluorescence qubit state detection.}
\end{figure}

In a successful entanglement event, the emitted photon passes through a polarization rotator ($\lambda/2$ waveplate) followed by a polarizing beam splitter [Fig.~\ref{fig:f2}] that directs the two polarization components to photon-counting photomultiplier tubes (PMTs), providing a measurement of the photonic qubit state.  Following the photon detection on either PMT, two phase-coherent microwave pulses are applied to the ion.  The first pulse performs a complete population transfer from the $|1,1\rangle$ ground state to the $|0,0\rangle$ ground state [Fig.~\ref{fig:f1}(b)], thus transferring the atomic qubit to the states $|1,0\rangle\equiv|0\rangle_S$ and $|0,0\rangle\equiv|\tilde{1}\rangle_S$, which is required for atomic qubit state detection as described below.  The second pulse, resonant with the $|0,0\rangle \leftrightarrow |1,0\rangle$ transition, rotates the atomic qubit by any amount in the Bloch sphere R$_S(\theta_{S},\phi_{S}$) determined by the pulse length and phase [Fig.~\ref{fig:f1}(c)].  Following the microwave rotations, the atomic qubit state is measured to better than 95\% efficiency via standard trapped-ion flourescence techniques \cite{blatt:1988, deslauriers:2004}: if the ion is in the $F=1$ manifold, a 125 $\mu$s, $\sigma^+$-polarized laser pulse resonant with the excited state drives a cycling transition, resulting in a high flourescence rate from the ion, whereas if the ion is in the $F=0$ state ($|\tilde{1}\rangle_S$), the laser frequency is far detuned and results in nearly no flourescence detected by the PMTs.

The use of two microwave pulses is not only important for subsequent atomic qubit detection, but also simplifies the phase-locking of atomic and photonic qubit rotations.  In order to reliably rotate the atomic qubit with respect to the photonic qubit, it is important to control the microwave phase with respect to the arrival time of the photon, which occurs randomly within a 50 ns window.  For example, if the atomic qubit begins in the initial state $(|0\rangle_S + |\tilde{1}\rangle_S)/\sqrt{2}$ and is rotated directly with a single microwave pulse, R$_S(\theta_S$,$\phi_S$), the final state measurement would depend on the absolute phase of the microwave source, with a probability of measuring $|0\rangle_S$ of $P(|0\rangle_S) = (1-\text{cos}\phi_S\text{sin}\theta_S)/2$.  However, if before the qubit rotation pulse, the population of one of the qubit states is first completely transfered to another state (i.e. $|1\rangle_S\rightarrow|\tilde{1}\rangle_S$ via a R$_S(\tilde{\theta}_S=\pi$,$\tilde{\phi}_S$) transfer pulse), then the final state measurement depends only on the phase \textit{difference} between the two microwave pulses: $P(|0\rangle_S) = (1-\text{cos}(\phi_S-\tilde{\phi}_S)\text{sin}\theta_S)/2$.  Hence, the phase of the atomic qubit rotation can be easily controlled by setting the relative phase of the two microwave sources \cite{blinov:2004}.  

A complete measurement of the CHSH form of Bell inequality requires the accumulation of four correlations, with maximum violation occurring when one qubit is rotated by $\theta_A=0,\pi/2$ and the other by $\theta_B=\pi/4,3\pi/4$.  Here, two complete inequality measurements are taken by rotating the ion by $\theta_S=0,\pi/2$ ($\theta_S=\pi/4,3\pi/4$), and rotating the photon by $\theta_P=\pi/4,3\pi/4$ ($\theta_P=0,\pi/2$).  Each correlation measurement consists of approximately 2000 successful entanglement events and takes around 20 minutes, requiring about 80 minutes for the complete Bell inequality measurement.  From these correlation measurements, the Bell signals are $B=2.203\pm0.028$ ($B=2.218\pm0.028$), where the uncertainties are statistical (Table \ref{table1}).  One possible source of systematic error considered is the unequal efficiencies of the photon detectors.  In order to give the proper weight to the data collected on each PMT, each correlation measurement consists of two runs, in which the role of the two PMTs are reversed via a 45 degree rotation of the $\lambda/2$ waveplate.

\begin{table}
\caption{\label{table1}Result of the two Bell inequality experiments}
\begin{ruledtabular}
\begin{tabular}{lcr}
$\theta_S$ & $\theta_P$ & q($\theta_S$,$\theta_P$)\\
\hline
$0$ & $\pi/4$ & 0.558\\
$0$ & $3\pi/4$ & -0.519\\
$\pi/2$ & $\pi/4$ & 0.513\\
$\pi/2$ & $3\pi/4$ & 0.613\\
\end{tabular}
$B(\theta_{S1}$=$0,\theta_{S2}$=$\pi/2;\theta_{P1}$=$\pi/4,\theta_{P2}$=$3\pi/4)=2.203\pm0.028$\\
\begin{tabular}{lcr}
$\theta_S$ & $\theta_P$ & q($\theta_S$,$\theta_P$)\\
\hline
$\pi/4$ & $0$ & 0.636\\
$\pi/4$ & $\pi/2$ & 0.461\\
$3\pi/4$ & $0$ & -0.516\\
$3\pi/4$ & $\pi/2$ & 0.605\\
\end{tabular}
$B(\theta_{P1}$=$0,\theta_{P2}$=$\pi/2;\theta_{S1}$=$\pi/4,\theta_{S2}$=$3\pi/4)=2.218\pm0.028$
\end{ruledtabular}
\end{table}

Upon emission of the photon, the two qubits are ideally in the entangled state $|\Psi\rangle_{ideal} = (|0\rangle_S |0\rangle_P+|1\rangle_S |1\rangle_P)/\sqrt{2}$, while the actual prepared state, represented by the density matrix $\rho$, has a fidelity of F = $_{ideal}$$\langle\Psi|\rho|\Psi\rangle_{ideal}\cong0.87$ \cite{blinov:2004}.  Depending on the particular decomposition of the density matrix, this should produce a Bell signal between 2.09 and 2.46.  Hence, our measured values agree well with the predictions of quantum mechanics and violate the Bell inequality by greater than seven standard deviations. 

While these results are in good agreement with quantum mechanics, neither the detection nor the locality loopholes are closed.  The probabilistic nature of photon detection leaves open the detection loophole, as the results rely on the detected events representing a ``fair sample'' of the entire ensemble.  The locality loophole is not closed since the photon's polarization rotation and detection take place approximately 1.1 meters away from the atom, and the detection of the atomic qubit takes 125 $\mu$s, falling well within the backward lightcone of the photonic detection event.  On the other hand, the detection and locality loopholes have been previously closed in two separate experiments involving pairs of entangled ions \cite{rowe:2001} and entangled photons \cite{weihs:1998}, respectively.  Yet, despite a number of proposals \cite{kwiat:1994, huelga:1995, fry:1995, simon:2003}, no experiment to date has simultaneously closed both loopholes.  

The experiment reported here, however, may provide a path toward a loophole-free Bell inequality test involving pairs of remotely-entangled ions \cite{simon:2003}.  This begins by first entangling two ion-photon pairs simultaneously in distant locations, as above.  Next, each emitted photon is directed (via fibers) to an intermediate location where a partial Bell state analysis is performed \cite{braunstein:1995, mattle:1996}, thus projecting the remotely-located ions into a known entangled state.  This entangled-ion pair is the starting place for a loophole-free Bell inequality test, which is completed by independently rotating each ion on the Bloch sphere, followed by qubit state detections.  By reducing the time needed for the atomic qubit rotation and detection to 50 $\mu$s, a separation of 15 km is sufficient to satisfy the time constraints of the locality loophole.  With this shorter detection time, sufficiently high atomic qubit detection efficiency is still possible to close the detection loophole, however, transmission of the ultraviolet photons over the necessary 7.5 km to the analyzer would be difficult with current technologies.  One could also perform the experiment with a more suitable photon color via frequency conversion \cite{james:2002} or by using a different ion species.  Additionally, one could use a quantum repeater method, entailing intermediately-located ion traps, thereby decreasing the distance any one photon needs to travel \cite{briegel:1998, dur:1999, duan:2004}.  Because of the long coherence and storage times of trapped ions, a network of remotely entangled ions would be not only useful for a loophole-free Bell inequality test, but also for scalable quantum computation and communication and as the building blocks for a quantum communication network. 

In conclusion, we report the first experimental observation of a Bell inequality violation between particles of different species, represented by a single atom and a single photon.  In addition, we have laid the groundwork for a loophole-free Bell test and a robust scalable quantum computation and communication architecture.

\begin{acknowledgments}
We wish to thank L.- M. Duan for his useful advice, C. Simon for useful discussions, and R. Miller for assistance in building the trap apparatus.  This work was supported by the U.S. National Security Agency and Advanced Research and Development Activity under Army Research Office contract DAAD19-01-1-0667 and the National Science Foundation Information Technology Research Program.  
\end{acknowledgments}


\begin{thebibliography}{28}
\bibitem[{cor()}]{correlation:footnote}
\bibinfo{note}{This is equivalent in form to the expectation value of the
  parity operator, $\langle\sigma_{z}^{(A)} \sigma_{z}^{(B)}\rangle$.}
\end{thebibliography}

\end{document}